\documentclass[aps,prmaterials,superscriptaddress,twocolumn]{revtex4}
\usepackage{amssymb,mathrsfs}

\usepackage{color}

\usepackage{graphicx}
\usepackage{amsmath}
\usepackage{epsfig}
\usepackage{times}
\usepackage{caption}
\usepackage{float}
\usepackage{pdflscape}
\usepackage{rotating}
\DeclareCaptionLabelSeparator{dot}{. }
\usepackage{titlesec}
\begin{document} 
\renewcommand{\figurename}{FIG.}

\bibliographystyle{apsrev}

\title{Unexpected Termination Switching and Polarity Compensation in LaAlO$_{3}$/SrTiO$_{3}$ heterostructures} 

\author{Guneeta Singh-Bhalla}
\affiliation{Department of Physics, University of California, Berkeley, California 94720, USA}
\affiliation{Materials Science Division, Lawrence Berkeley National Laboratory, Berkeley, California 94720, USA}

\author{Pim B. Rossen}
\affiliation{Department of Materials Science and Engineering, University of California, Berkeley, California 94720, USA}

\author{Gunnar K. P\'alsson}
\altaffiliation[Current address: ]{Department of Physics and Astronomy, Uppsala University, Uppsala, 751 20, Sweden}
\affiliation{Materials Science Division, Lawrence Berkeley National Laboratory, Berkeley, California 94720, USA}
\affiliation{Department of Physics, University of California, Davis, CA 95616, USA}

\author{Jaganatha S. Suresha}
\affiliation{National Center for Electron Microscopy, Lawrence Berkeley National Laboratory, Berkeley, California 94720, USA}

\author{Di Yi}
\altaffiliation[Current address: ]{Department of Applied Physics, Stanford University, Stanford, CA 94305, USA}
\affiliation{Department of Materials Science and Engineering, University of California, Berkeley, California 94720, USA}

\author{Abhigyan Dasgupta}
\affiliation{Department of Physics, University of California, Berkeley, California 94720, USA}

\author{David Doenning}
\affiliation{Department of Earth and Environmental Sciences and Center of Nanoscience (CENS),
University of M\"unich, DE-80333 M\"unich, Germany}

\author{Victor G. Ruiz}
\affiliation{Department of Earth and Environmental Sciences and Center of Nanoscience (CENS),
University of M\"unich, DE-80333 M\"unich, Germany}

\author{Ajay K. Yadav}
\affiliation{Department of Materials Science and Engineering, University of California, Berkeley, California 94720, USA}

\author{Morgan Trassin}
\altaffiliation[Current address: ]{Department of Materials, ETH Zurich, Vladimir-Prelog-Weg 4, 8093 Zurich, Switzerland}
\affiliation{Department of Materials Science and Engineering, University of California, Berkeley, California 94720, USA}

\author{John T. Heron}
\altaffiliation[Current address: ]{Department of Materials Science and Engineering, University of Michigan, Ann Arbor, Michigan 48109, USA}
\affiliation{Department of Materials Science and Engineering, University of California, Berkeley, California 94720, USA}

\author{Charles S. Fadley}
\affiliation{Materials Science Division, Lawrence Berkeley National Laboratory, Berkeley, California 94720, USA}
\affiliation{Department of Physics, University of California, Davis, CA 95616, USA}

\author{Rossitza Pentcheva}
\affiliation{Department of Physics and Center for Nanointegration  (CENIDE), University of Duisburg-Essen, 47057 Duisburg, Germany}

\author{Jayakanth Ravichandran}
\affiliation{Mork Family Department of Chemical Engineering and Materials Science, University of Southern California, Los Angeles, CA 90089, USA}
\email[Corresponding author:~]{jayakanr@usc.edu}

\author{Ramamoorthy Ramesh}
\email[Corresponding author:~]{rramesh@berkeley.edu}
\affiliation{Department of Physics, University of California, Berkeley, California 94720, USA}
\affiliation{Materials Science Division, Lawrence Berkeley National Laboratory, Berkeley, California 94720, USA}
\affiliation{Department of Materials Science and Engineering, University of California, Berkeley, California 94720, USA}

\date{\today}

\def\la{LaAlO$_3$ }
\def\lan{LaAlO$_3$}
\def\sr{SrTiO$_3$ }
\def\srn{SrTiO$_3$}
\def\al{AlO$_2^-$ }
\def\alm{AlO$_x^-$ }
\def\ti{TiO$_2$ }
\def\p{$p$-type }
\def\n{$n$-type }





\begin{abstract}
Polar crystals composed of charged ionic planes cannot exist in nature without acquiring surface changes to balance an ever-growing dipole. The necessary changes can manifest structurally or electronically.  An electronic asymetry has long been observed in the \lan/\sr system. Electron accumulation is observed near the \lan/TiO$_2$-SrTiO$_3$ interface, while the LaAlO$_3$/SrO-SrTiO$_3$ stack is insulating.  Here, we observe evidence for an asymmetry in the surface chemical termination for nominally stoichiometric LaAlO$_3$ films in contact with the two different surface layers of SrTiO$_3$ crystals, TiO$_2$ and SrO. Using several element specific probes, we find that the surface termination of \la remains AlO$_{2}$ irrespective of the starting termination of \sr substrate surface. We use a combination of cross-plane tunneling measurements and first principles calculations to understand the effects of this unexpected termination on band alignments and polarity compensation of LaAlO$_3$/SrTiO$_3$ heterostructures. An asymmetry in \la polarity compensation and resulting electronic properties will fundamentally limit atomic level control of oxide heterostructures. \end{abstract}

\maketitle 



\setlength{\epsfxsize}{2\columnwidth}
\begin{figure*}[t]
\includegraphics[width=2\columnwidth]{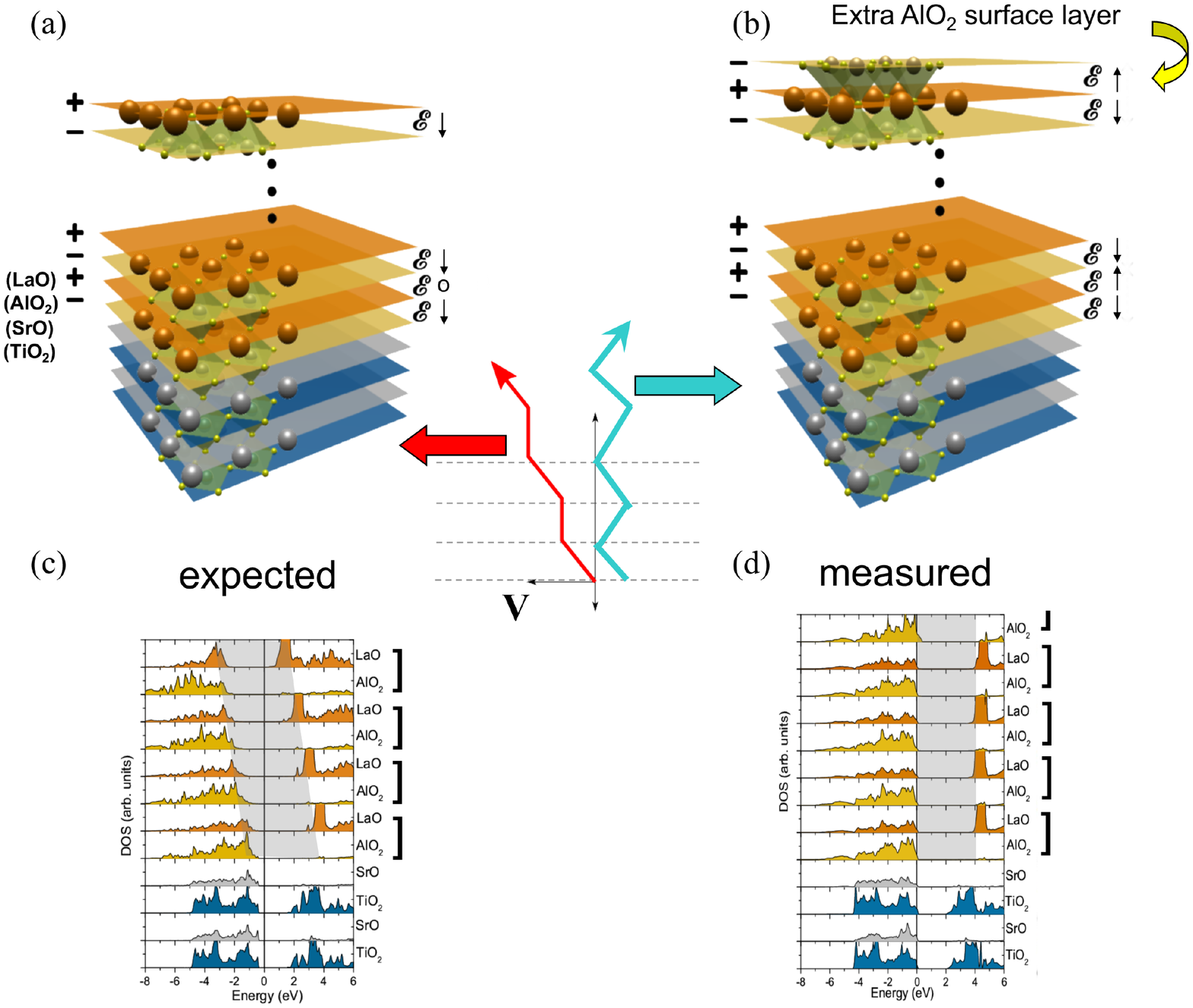}
\caption{A comparison of the theoretically expected and experimentally observed \p \lan/\sr heterostructure, both with and without an additional ionic layer that cancels \lan's diverging dipole.  (a) The schematic depicts the expected atomic layer stacking for a \p \lan/\sr heterostructure.  \sr terminates with SrO (grey) at the interface, in direct contact with \al (yellow).  \la terminates at the LaO$^+$ (orange) surface layer.  Vertical ellipses convey an arbitrary \la thickness.  The net charge in the ionic limit is labeled to the left of the  \la stack, while the net electric field ($\mathscr{E}$) is depicted on the right using arrows.  The center inset depicts the resulting potential which grows with thickness (blue) for the expected \p structure.  (b) The \p \la stack with an additional \alm (ideally AlO$_2^-$) layer as confirmed by measurements is shown. Neighboring layers have an equal and opposite $\mathscr{E}$, resulting in no net potential (red arrow, center inset). Hence the extra surface \alm layer cancels the dipole.  (c) The layer resolved DOS (LDOS) calculated using DFT are shown for a \p 4~u.c. thick \la stack, as depicted in (A), with the LaO$^+$ top surface. The presence of $\mathscr{E}$ is reflected in shifts of the O $2p$ and La $4f$ bands.  (d) LDOS of the same system as in (c) adding an extra \al layer diminishes $\mathscr{E}$.}
\end{figure*}

When cleaved, oxide crystals composed of charged ionic planes may become highly unstable depending on the crystallographic orientation and the associated charges of the exposured surface.  An electric dipole moment that is inherent to the layered structure can emerge, giving rise to a potential that diverges with thickness.~\cite{Tasker1979} Such polar instabilities in nature may be compensated through complicated surface reconstruction processes that include rumpling of the surface atoms, ad-atom absorption, stoichiometric changes or screening charge accumulation.~\cite{Hesper2000,Goniakowski2008,noguera2008a}  An additional degree of complexity arises when polar crystals are directly in contact with other materials and band alignments influence the reconstruction process.~\cite{noguera2008a,Gu2009}  For instance, a curious asymmetry is observed when polar LaAlO$_3$ films are grown on SrTiO${_3}$ crystals with either an SrO or TiO$_2$ surface layer.  An electron gas appears at the ``$n$-type" SrTiO${_3}$-TiO$_2$/\la interface for \la thicker than 4-6 unit cells (u.c.), but the ``$p$-type" SrTiO${_3}$-SrO/LaAlO$_3$ interface is insulating.~\cite{ohtomo2004,Thiel2006,huijben2006}  Accumulation of free charge at the \n interface is thought to play a role in screening the dipole across \lan.~\cite{nakagawa2006,singhbhalla2011,xie2011} Analogously, a dipole is also expected to form across \la for the \p system requiring positive charge at the interface for screening, but experimentally the interface is insulating.~\cite{ohtomo2004,nakagawa2006}  One study suggests that localized oxygen vacancies at the \p interface may screen the dipole, although the work focused on multilayers, which differ from the bi-layer SrTiO${_3}$/LaAlO$_3$ heterostructure in both growth dynamics and band alignments.~\cite{nakagawa2006}  Furthermore, two core level photoemission results show evidence of finite but opposite polar fields across \la for the \p system, which are smaller than the ideal field for the \n interface.~\cite{segal2009,Takizawa2011} Hence the origin of the asymmetrical electric properties for the two heterostructures has heretofore remained largely unresolved.  A fundamental understanding of the polarization compensation mechanisms in model oxide heterostructures such as SrTiO${_3}$/LaAlO$_3$ is crucial for developing controlled interfaces and devices for oxide electronics.~\cite{noguera2008a}

Our observations for the \p heterostructure reveal an unexpected chemical change on the top LaAlO$_3$ surface, which may carry implications for electronic properties of \lan/\sr system.  For LaAlO$_3$ films grown on SrTiO$_3$ (001) single crystals, we expect to obtain \srn-TiO$_2$-SrO-TiO$_2$/LaO-AlO$_2$-LaO-AlO$_2$- $\cdots$ -LaO-AlO$_2$ for the $n$-type heterostructure and \srn-TiO$_2$-SrO/AlO$_2$-LaO-AlO$_2$-LaO $\cdots$ -AlO$_2$-LaO for the $p$-type structure (Fig.\ 1(a)). Assuming formal valencies as labeled in the Figs.\ 1(a) and 1(b), and using a parallel plate capacitor approximation, a net internal electric field ($\mathscr{E}$) will appear between every other layer as depicted by arrows.~\cite{Tasker1979,nakagawa2006}  Hence, a diverging dipole requiring surface reconstructions should be present for both heterostructures, albeit with opposite polarity (see red curve Fig.\ 1 inset for the \p system, and Figs.\ S1 and S2 for the \n system). In fact, theoretical calculations suggest that the polarity induced defects mechanisms could explain the electronic properties of polar heterostructures such as \lan/\sr system.~\cite{Yu:2014hx}  

In this work, we use a combination of three element specific probes such as Time of flight - ion scattering and recoil spectroscopy (TOF-ISARS), angle resolved x-ray photoelectron spectroscopy (AR-XPS), and high resolution scanning transmission electron microscopy (STEM) combined with cross-plane transport measurements and first principles calculations to probe the surface composition and its potential effects on polarity of \la for both heterostructures. Although formally we expect an LaO$^+$ surface layer for the \p heterstructure and \al for the \n heterostructure, using AR-XPS and TOF-ISARS, we measure instead an \alm surface layer for both systems. STEM studies confirm the findings, and moreover, reveal that the \p structure $acquires$ an extra AlO$_x^-$ layer, i.e.\ an extra half unit cell during growth, as depicted in Fig.\ 1(b). Measurements of tunnel current density across the \la layer help us better understand the potential implications for polarity compensation in the \p structure.
%

\setlength{\epsfxsize}{2\columnwidth}
\begin{figure*}[t]
\epsfbox{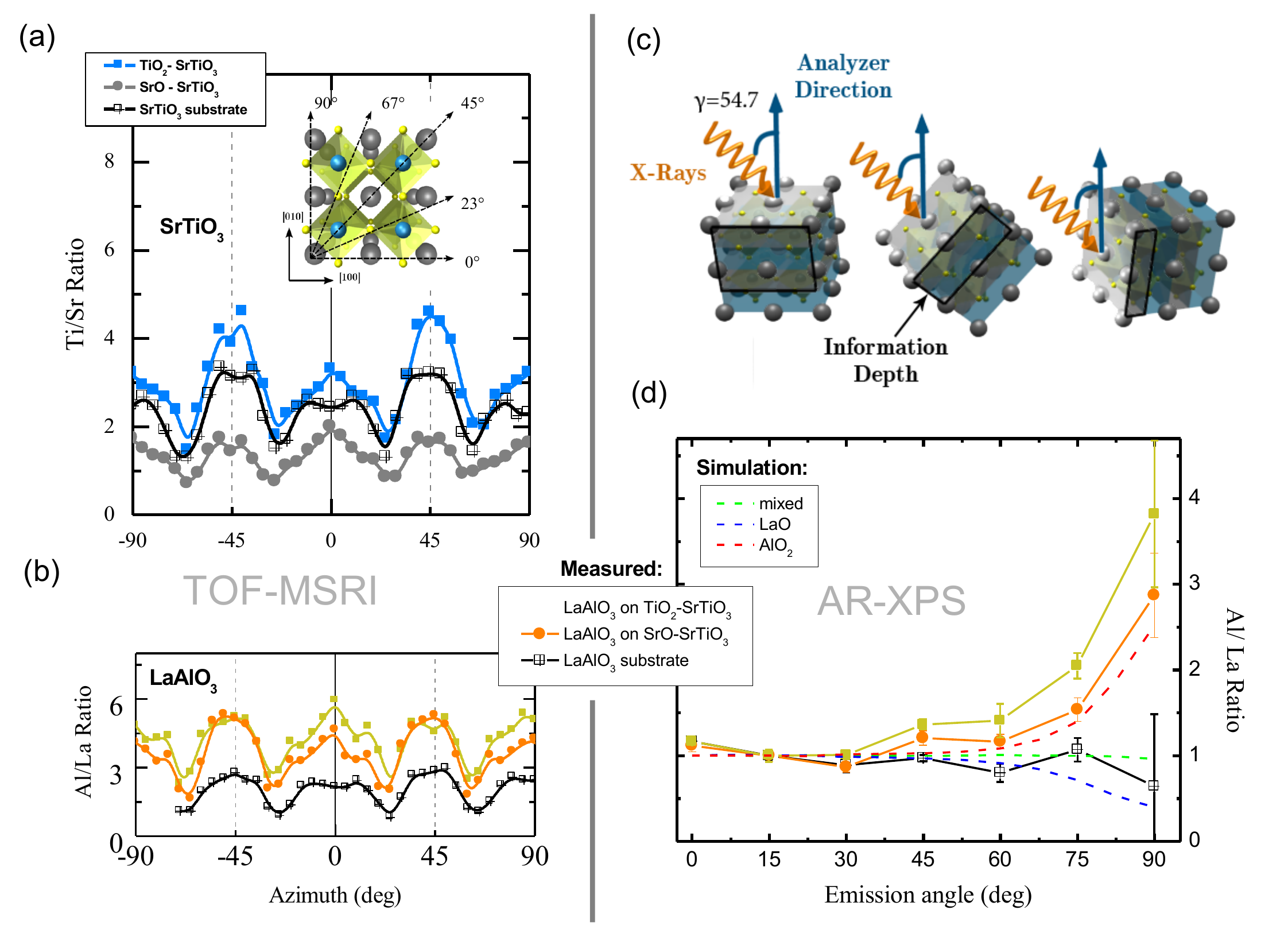}
\caption{ TOF-ISARS and AR-XPS measurements show an \alm top surface for \la in both the \n and \p heterostructure geometries.  (a) TOF-ISARS is used to confirm the surface termination of \sr substrates.  K$^+$ ions impinge the surface at an angle of 15$^\circ$ as the sample is rotated about the azimuth ($x$-axis).  The scattered and recoiled ions reflect the atomic structure and mass of the surface species.  At 45$^\circ$ when the A- and B-site atoms shadow one another, we see an enhanced ratio of Ti/Sr at 0$^\circ$ for the \ti terminated substrates compared to the SrO terminated substrates.  (b).  Using TOF-ISARS we find an Al$^+$ rich surface for \la grown on both the \ti (\n) and SrO (\p) \sr surfaces.  (c). The schematic depicts our AR-XPS set-up with a fixed angle $\gamma$ = 54.7$^\circ$ between the incoming x-rays and the analyzer.  The sample is rotated about an angle $\alpha$ (not shown) with respect to the analyzer. (d) AR-XPS measurements are compared to SESSA simulations for the various \la surface terminations as labeled.  Clearly, \lan's top surface terminates with an \alm layer for both the \p and \n geometries, in agreement with TOF-ISARS. Additional simulations comparing the measurement results to stoichiometric changes in sub-surface \la layers, as well as intermixing with the \sr elemental species are also considered.  The best agreement between the data and simulations are achieved by considering a stoichiometric \la layer terminated with an AlO$_x^-$ top layer.}
\end{figure*}

The details of the film growth and characterizations, and theoretical calculations are provided in Supplemental Materials. We begin the experimental studies by confirming the \ti and SrO surface terminations of \sr (001) oriented substrates using TOF-ISARS.  Two substrates of each termination were prepared using standard techniques.~\cite{kawasaki1994,rijnders2004,yu2012}  Fig.\ 2(a) shows relative intensities of the Ti${^+}$/Sr${^+}$ ratios obtained from the mass spectroscopy of the recoiled ions (MSRI).  Single termination is revealed in the form of a clear asymmetry in the ratio of Ti${^+}$ to Sr${^+}$ ions at a measurement angle of 45${^\circ}$, along the [110] direction for the two terminations.  As depicted in the inset to Fig.\ 2(a), at this angle the Ti${^+}$ and Sr${^+}$ ions shadow one another, hence only the topmost layer of SrTiO${_3}$ is probed.~\cite{gozar2007,Kleibeuker2010,biswas2011}  At 0${^\circ}$, the incident beam is aligned along [100] and [010] directions equally along rows of Sr${^+}$ and Ti${^+}$ ions. The mixed termination of an untreated SrTiO${_3}$ substrate on the other hand reveals nearly equivalent ratios for the Ti${^+}$ and Sr${^+}$ ions at 45${^\circ}$ and 0${^\circ}$.  

Following \la deposition, the surface was probed \textit{in situ} using the same technique for two of each \p and \n heterostructures.  Fig.\ 2(b) compares LaAlO$_3$ films grown on the two different SrTiO${_3}$ terminations along with a mixed termination bulk \la crystal with (001) orientation.  As noted above, we expect an LaO$^+$ surface layer for \la in the \p geometry and \al for the \n geometry.  Instead we find that \la in both the \n and \p geometries displays the same Al${^+}$/La${^+}$ ratio, indicating a surface layer rich in Al$^+$.  We note a slight decrease in the Al${^+}$/La${^+}$ ratio at zero degrees, indicating a slightly  Al${^+}$ poor surface for the $p$-type sample, as compared to the $n$-type sample.  
%

AR-XPS measurements of \la in each heterostructure corroborate the TOF-ISARS findings, clearly revealing an \alm termination for both types of interfaces.  By varying the angle of electron emission, $\alpha$, relative to the analyzer (depicted in Fig.\ 2(c)), the mean electron escape depth, ${\Lambda(\alpha)=\Lambda}$cos($\alpha$), is varied.  If an element is closer to the surface, its relative intensity will be enhanced as $\alpha$ is increased.   Fig.\ 2(d) compares the measured ratio between the areas of the Al-2s and La-4d peaks with simulations using the commercially available SESSA software package, for both heterostructures as well as the mixed termination LaAlO$_3$ crystal.~\cite{Smekel2005,SESSA} The ratio increases with $\alpha$ for both heterostructures but remains roughly constant for the mixed termination substrate.  This is consistent with SESSA simulations for an AlO$_2$ (maroon curve) and mixed \la termination (green curve) respectively. The results were reproduced for two additional copies of the samples on two different spectrometers.  LaAlO$_3$ in the $n$-type geometry terminates with the expected \alm surface layer.  If the \p \la surface were terminated with the expected LaO$^+$ layer one would instead expect a decrease in the Al-2s/La-4d ratio with increasing $\alpha$ (Fig.\ 2(d), blue curve).  However, in agreement with the TOF-ISARS results, we find that just as for the \n system, \la in the \p system also terminates with an \alm surface layer.  Furthermore, we find a slightly reduced ratio of Al-2s/La-4d ratio for the \p structure than for the \n structure, again in agreement with the TOF-ISARS results. The two surface probes thus reveal consistent results. In the Supplementary Section, the measured data are further compared to SESSA simulations for a variety of scenarios including off-stoichiometric films or surface layers with La${^+}$/Al${^+}$ ratios above or below unity, as well as Al${^+}$ rich and Al${^+}$ poor surfaces.  As seen in the Supplementary Section (Fig. S3), the best agreement between the SESSA simulations and measured data was achieved for a stoichiomentric \la film grown on either substrate termination of \sr with a top surface of \al in both cases.  

To learn about the atomic structure of the entire \srn/\la stack and allow the visualization of its layer arrangement, high-resolution STEM imaging was carried out using the aberration-corrected TEAM 0.5 microscope. Cross-sectional STEM imaging of the layered architecture for both the \p (top) and \n (bottom) heterostructures is shown in Fig.\ 3(a), suggesting the presence of atomically abrupt, epitaxial interfaces between \la and \srn. The images were acquired along the perovskite pseudocubic direction (surface normal) and exhibit atomic columns with two distinct intensities:  the La and Sr atomic columns appear brighter than the Ti and Al columns.   Fig.\ 3(b) shows high-angle annular dark field (HAADF) intensity profiles along the line displayed in the corresponding STEM images (Fig.\ 3(a)). The intensity profile clearly shows six and seven \al layers for nominally six unit cells of \la film in the \n and \p type heterstructures. The extra layer in the \p heterostructure is highlighted in the image. From the HAADF intensity profiles, and HAADF-STEM images, the interface atomic configurations and the terminations are deduced to be \srn-TiO$_2$/LaO$^+$-\al- $\cdots$-\alm for the \n heterostructure and \srn-SrO/\al-LaO$^+$- $\cdots$ -\alm for the \p structure.

\setlength{\epsfxsize}{1\columnwidth}
\begin{figure}[t]
\epsfbox{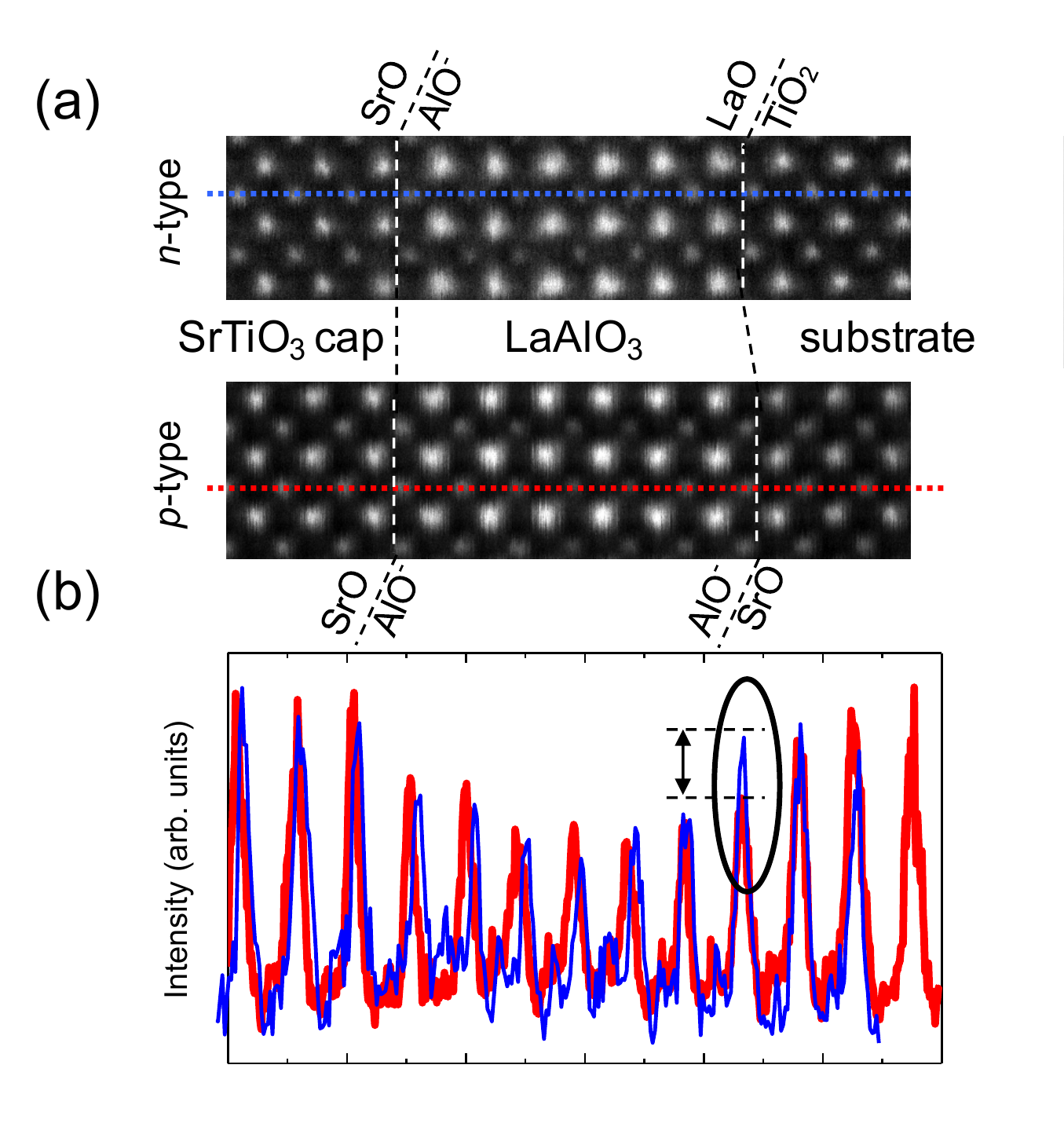}
\caption{HAADF-STEM images confirm \sr substrate terminations (\ti and SrO), and reveal the \alm \la surface termination for both \n and \p heterostructures.  (a) Cross-sectional STEM images for 6~u.c.\ thick \n and \p heterostructures, capped with a 10~u.c.\ thick \sr layer are shown in the top and bottom panel respectively.  The lines shown correspond with line scans in (b).  The \p image shows one extra half u.c. (b) HAADF intensity profiles for the \n (blue) and \p (red) heterostructure cross sections as shown in (a). The additional peak in the \p profile, circled for clarity, highlights the additional Al$^{3+}$ rich atomic layer for the \p heterostructure.}
\end{figure}

%

The HAADF-STEM results corroborate the TOF-ISARS and AR-XPS results for the \alm surface termination for \la in both the \p and \n heterostructures, in addition to the \sr substrate termination. Most importantly, we learn that \textit{an extra} \alm layer is present for the \p interface.  In this regard, when considering formal valencies and approximating the ionic charge for each layer within \lan, an important consequence of the observed surface termination is the existence of a polarity built into \lan~for the $n$-type heterostructure (Figs.\ 1(a) ), but ionically compensated for the $p$-type system by the extra \alm layer (Fig.\ 1(b)).  This key difference between the dipoles across \lan  for the two stacking geometries can be verified using density functional theory (DFT) as shown in Figs.\ 1C, 1D and Fig.\ S1.  For a 4 u.c.\ thick \la film within the expected \p geometry, Fig.\ 1C shows the presence of an internal electric field, $\mathscr{E}$ reflected in shifts of the O $2p$ and La $4f$ bands in subsequent layers to lower energies.  The internal electric field is partially screened by lattice polarization that are of opposite sign to the ones observed for the $n$-type \lan/\sr system as shown in Fig.\ S2.~\cite{pentcheva2008,pentcheva2009}  Calculations for a $p$-type system where an additional AlO$_2$ layer is added on top of the 4 u.c.\ \la film are shown in Fig.\ 1D. In contrast to the LaO$^+$ terminated system, the layer resolved density of states (DOS) shows that $\mathscr{E}$ is now canceled and the system is insulating except for some holes in the surface layer that are likely to be compensated by oxygen vacancies. With the presence of an additional \al layer on the \la top surface, we find that there is no measureable lattice polarization. Hence, within the DFT simulation, the extra \al layer screens the dipole for the \p system.

While the mechanism for accumlating an addition \al layer is unclear and beyond the scope of the present study, we probe the influence it may have on \la polarity.  A common method for measuring the electronic signatures of a built-in potential across polar dielectrics entails measuring the tunneling current density ($J$) across the dielectric layer as a function of thickness, $d$, within a parallel plate capacitor geometry, i.e., a metal-insulator-metal (MIM) tunnel junction. For a typical insulator in a MIM configuration, when a set bias voltage, $V$, is applied across the junction, $J$ decreases exponentially with increasing $d$, thus $J$ $\propto$ $e^{-d}$.  If however, there is a built-in potential ${V_{bi}}$ present across the dielectric, the applied $V$ will be offset as ${V_{bi}}$ grows with thickness.  As the thickness of a polar dielectric such as \la varies, any unscreened potential will grow with $d$ bending the bands in the dielectric until the valence band crosses the Fermi level at a critical thickness, $d^{\mathrm{cr}}$, and results in a sudden large increase of the overall tunneling current density, $J$.  Such an increase in $J$ by orders of magnitude was previously observed at a critical thickness for \n structures.~\cite{simon2009, singhbhalla2011} Hence the potential across a polar dielectric can be modulated by changing $d$ to affect  ${V_{bi}}$, or by modulating the applied bias, $V$. This phenemonon has previous been observed in wide band gap III-V nitrides and also ferroelectric insulators.~\cite{Bykhovski1995, simon2009, wetzel2000}  

Such $J$ vs. $d$ analysis has previously been carried out in great detail for the \n \lan/\sr system, in direct analogy to the wide-bandgap polar III-V nitrides.~\cite{simon2009, singhbhalla2011}  Intriguingly, a $d_{\mathrm{LAO}}^{\mathrm{cr}}$ was identified at which a sudden increase in $J$ by orders of magnitude coincided with an alignment between the valence band of \la and the \sr conduction band, from which a value of 80.1~meV/\AA\ was extracted for the polar field across \la in the \n \lan/\sr geometry. For the \p structures in the present configuration, the electrostatic arguments depicted in Fig.\ 1 predict no ${V_{bi}}$ across \lan, so that $J$ $\propto$ $e^{-d}$ as for a typical MIM junction.  Hence, in stark contrast to the polarized \la in the \n structure, no sudden increase in $J$ would be expected at a critical thickness for the unpolarized \la in the \p structure.

\setlength{\epsfxsize}{1\columnwidth}
\begin{figure}[t]
\epsfbox{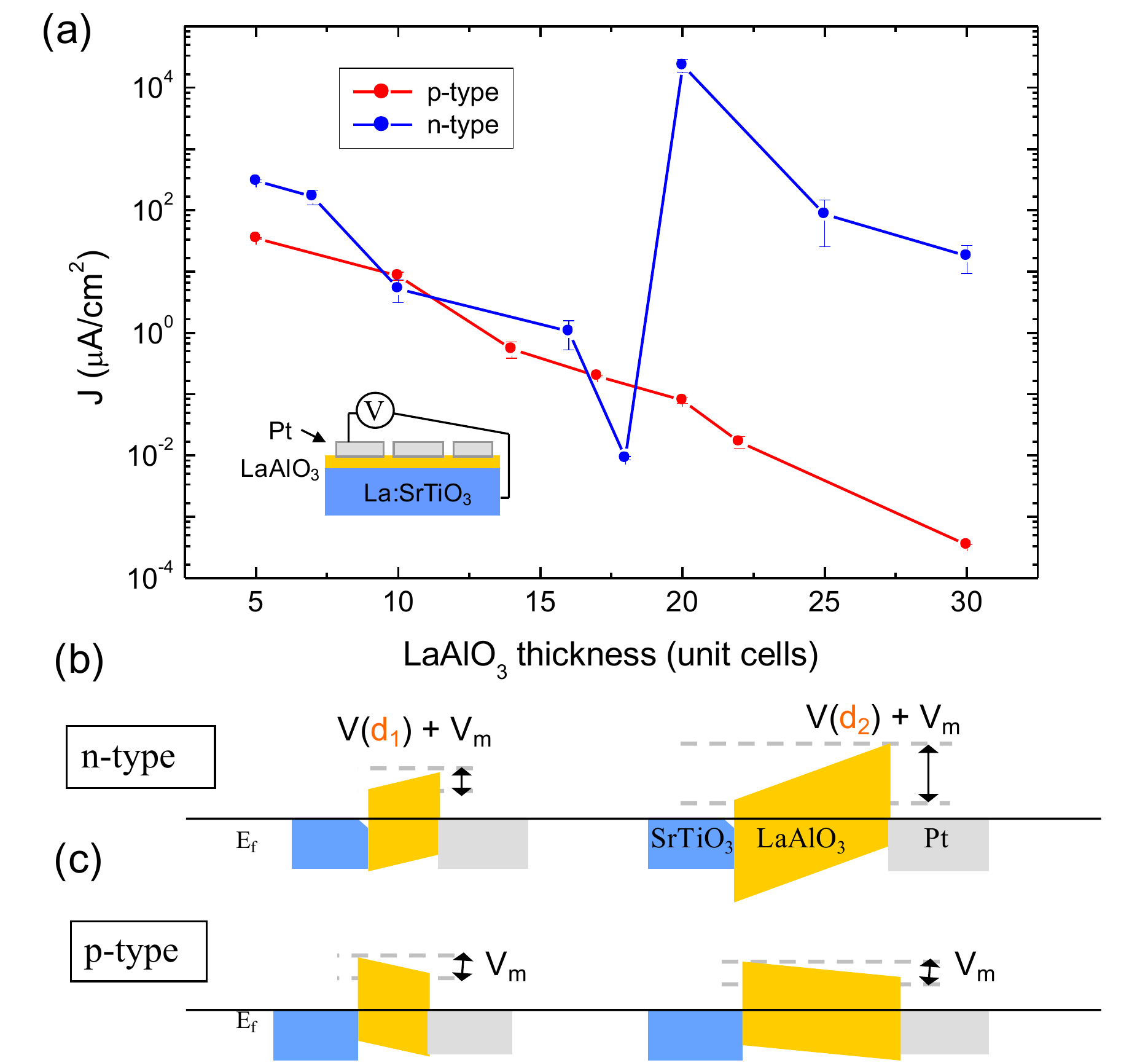}
\caption{Tunneling measurements reveal the presence of a dipole across \la in the \n geometry but not the \p geometry.  (a)  Room temperature measurements of tunneling current, $J$, as a function of barrier thickness $d_{\mathrm{LAO}}$, across Pt/\lan/La:\sr junctions is shown in log scale at $V$ = -0.1~V (\n) and 0.1~V (\p).  As expected, there is an exponential decrease in current up to $d_{\mathrm{LAO}}$ = 20~u.c.\ for both \p and \n junctions.  However, a sudden increase in $J$ is observed above 20~u.c.\ as previously observed~\cite{singhbhalla2011} in the presence of a dipole.  $J$ across the \p junction however continues to exponentially decrease with increasing thickness, revealing no signatures of a dipole. (b) and (c) show schematics depicting possible \n and \p Pt/\lan/La:\sr junction band diagrams with a band misalignment potential $V_m$ that remains constant with thickness and ${V(d_{\mathrm{LAO}})}$, the potential arising from an unscreened dipole that grows with thickness.}
\end{figure}

In order to verify this hypothesis, we measured the tunnel current density ($J$) between SrTiO${_3}$ and a Pt electrode evaporated on the LaAlO$_3$ surface (inset of Fig.\ 4(a)), as a function of \la thickness (${d_{\mathrm{LAO}}}$).  For the present study, $J$, was measured as a function of applied bias, $V$, between the Pt electrode and 0.01\% wt La$^+$ doped \sr substrates, which provide a conducting bottom electrode especially for the insulating $p$-type interface. Four tunnel junctions on each of seven \n and \p samples of varying thicknesses were measured.  The doping level of the substrates results in a bulk carrier density below that of the induced electron gas, ensuring that an accumulation of free charge is still required to screen the LaAlO$_3$ built-in field.  Fig.\ 4(a) shows $|J|$ vs. $d_{\mathrm{LAO}}$ as measured for $V = 0.1$~V for the \p heterstructures and $V = -0.1$~V for the \n heterostructure.  We choose the opposite polarity of $V$ for the \n and \p systems because a built-in field of opposite polarity would be expected for each (for electrons tunneling from the Pt electrode to the electron gas and vice versa, respectively).~\cite{simon2009, singhbhalla2011}   While a sudden increase of $J$, nearly seven orders of magnitude, is found at $d_{\mathrm{LAO}}^{\mathrm{cr}}$ = 20~u.c.\ for the $n$-type heterostructure, as previously observed in Ref.~\cite{singhbhalla2011}, no such increase in $J$ is found for the $p$-type heterostructure. As expected, \p system behaves as a typical MIM tunnel junction, where $J$ $\propto$ $e^{d_{\mathrm{LAO}}}$. Relaxation of the LaAlO$_3$ films with thickness~\cite{Cancellieri2011} cannot alone explain the sudden increase in $J$ measured across the $n$-type heterostructure at 20~u.c.\ since the same would also be expected for the $p$-type heterostructure, but is not observed. We also show reciprocal space maps of the \p and \n heterostructures (see Supplementary Section Fig.\ 5) to rule out the possibility of strain relaxation and/or other spurious mechanisms causing this large, abrupt change in the tunneling current density for the \n structure, but not the \p structure.  



The observed $|J|$ vs. $d_{\mathrm{LAO}}$ trend can be understood as follows.  Figs.\ 4(b) and 4(c) illustrate hypothetical band diagrams for two thicknesses of the \n and \p systems at $V = 0$.  The net ${V_{bi}}$ for a given \la thickness, $d_{\mathrm{LAO}}$ is the sum of net ionic potential at that thickness ${V_i(d_{\mathrm{LAO}})}$, plus the potential due to the band-bending resulting from work-function mismatch, ${V_m}$.  Hence ${{V_{bi}} = V_i(d_{\mathrm{LAO}}) + V{_m}}$.  While ${V_m}$ remains constant with thickness, ${V_i(d_{\mathrm{LAO}})}$ will grow with thickness in the presence of a finite \la dipole until the conduction and valence bands align, giving rise to a sudden increase in $J$.~\cite{singhbhalla2011}  Hence, in accord with previous results,~\cite{simon2009,singhbhalla2011} the blue curve in Fig.\ 4(a) implies a finite $V_i(d_{\mathrm{LAO}})$ for the \n system, while the red curve implies $V_i(d_{\mathrm{LAO}}) =$ 0 for the \p system, as sketched in Figs.\ 4(b) and 4(c).  Our results imply that while a dipole can be revealed by externally applying a potential across LaAlO$_3$ that perturbs the screening charge for the $n$-type heterostructure, as previously measured,~\cite{singhbhalla2011} for the $p$-type heterostructure evidence of an analogous dipole is not detected. Hence it is possible, as suggested by the first principles results above, that the \alm layer indeed cancels the \la polarity through an ionic (rather than electronic) reconstruction.  Our observations may help explain the lack of conductivity in \sr at the $p$-type interface, though further study by a variety of probes would be required for verification.

Thus, transport measurements suggest that no built-in polarity is present across \la in the \p configuration with the extra \al layer.  In light of the transport measurements combined with the discovery of the added \al layer, it is tempting to speculate that the extra layer forms as a result of energetics: i.e.\ it could be energetically more favorable for the system to accumulate an extra \alm top surface layer than to sustain a divergent dipole in the \p configuration.  To this end we note that our result is in agreement with a recent photoemission study where the core level shifts for the \p system did not qualitatively agree with theoretical predictions (similar to Figs.\ 1(a) and 1C), and the authors speculate that negative charge in some form must accumulate on the top surface of \lan.~\cite{Takizawa2011} Further studies, beyond the scope of the present work are needed understand the origin of the \al layer and it's influences on the \lan/\sr system.  For instance, it is possible that the additional \alm layer may accumulate immediately following growth by attracting residual Al$^{3+}$ atoms from the depostion atmosphere, or among other scenarios, it is also for instance possible that Al$^{3+}$ from the bulk of the film migrates to the surface to form the extra layer following deposition. As shown in the supplementary section, AR-XPS simulations rule out the latter scenario.  

In summary, we have shown compelling evidence for an unexpected top surface termination for \la grown on SrO terminated \srn, using a variety of element specific probes. Intriguingly the system acquires an additional half unit cell. Results from cross plane tunneling current measurements across \lan, and first principles calculations suggest that in the presence of the extra \al layer, polarity across the \p \lan/\sr is compensated and the system is in electrostatic equillibrium.  The observed asymmetry may indicate a basic limitation to bottom up, atomic scale design of material surfaces and interfaces, especially complex oxides, which are subject to various competing degrees of freedom. For instance, we have shown that control of \lan's top surface via the well known technique of \sr surface termination is non-trivial.  Essentially, a single atomic layer (SrO vs. TiO$_2$) can dramatically alter the way in which heterostructures balance various energy scales such as work function mismatch and built-in ionic potentials. Additional studies of polarity compensation and the critical length scales involved are needed for a variety of materials in order to develop a general understanding of the reconstruction processes and dynamics in polar heterostructures. 

The authors thank D. Schlom and D. Meier for insightful discussions and review of the manuscript. Electron microscopy was performed at the National Center for Electron Microscopy, Lawrence Berkeley National Laboratory. The AR-XPS were conducted at the Materials Science Division, Lawrence Berkeley National Laboratory. TOF-ISARS and tunneling measurements were conducted at UC Berkeley. The work was supported by the Office of Science, Basic Energy Sciences, Materials Sciences and Engineering Division of the US Department of Energy under contract DE-AC02-05CH11231. G.K.P. acknowledges the International Union for Vacuum Science, Technique and Applications and the Swedish Research Council for Financial Support, with partial support from the Army Research Office Multi-University Research Grant W911-NF-09-1-0398. J. R acknowledges support from the Air Force Office of Scientific Research (AFOSR) under Grant FA9550-16-1-0335. R.P. and D.D. acknowledge funding by the DFG within SFB/TR80 project C03/G03 and computational time at the Leibniz Rechenzentrum, project pr87ro.


\begin{thebibliography}{28}
\expandafter\ifx\csname natexlab\endcsname\relax\def\natexlab#1{#1}\fi
\expandafter\ifx\csname bibnamefont\endcsname\relax
  \def\bibnamefont#1{#1}\fi
\expandafter\ifx\csname bibfnamefont\endcsname\relax
  \def\bibfnamefont#1{#1}\fi
\expandafter\ifx\csname citenamefont\endcsname\relax
  \def\citenamefont#1{#1}\fi
\expandafter\ifx\csname url\endcsname\relax
  \def\url#1{\texttt{#1}}\fi
\expandafter\ifx\csname urlprefix\endcsname\relax\def\urlprefix{URL }\fi
\providecommand{\bibinfo}[2]{#2}
\providecommand{\eprint}[2][]{\url{#2}}

\bibitem[{\citenamefont{Tasker}(1979)}]{Tasker1979}
\bibinfo{author}{\bibfnamefont{P.~W.} \bibnamefont{Tasker}},
  \bibinfo{journal}{J. Phys. C: Sol. Stat. Phys.}
  \textbf{\bibinfo{volume}{12}}, \bibinfo{pages}{4977} (\bibinfo{year}{1979}).

\bibitem[{\citenamefont{Hesper et~al.}(2000)\citenamefont{Hesper, Tjeng,
  Heeres, and Sawatzky}}]{Hesper2000}
\bibinfo{author}{\bibfnamefont{R.}~\bibnamefont{Hesper}},
  \bibinfo{author}{\bibfnamefont{L.~H.} \bibnamefont{Tjeng}},
  \bibinfo{author}{\bibfnamefont{A.}~\bibnamefont{Heeres}}, \bibnamefont{and}
  \bibinfo{author}{\bibfnamefont{G.~A.} \bibnamefont{Sawatzky}},
  \bibinfo{journal}{Phys. Rev. B} \textbf{\bibinfo{volume}{62}},
  \bibinfo{pages}{16046} (\bibinfo{year}{2000}).

\bibitem[{\citenamefont{Goniakowski et~al.}(2008)\citenamefont{Goniakowski,
  Finocchi, and Noguera}}]{Goniakowski2008}
\bibinfo{author}{\bibfnamefont{J.}~\bibnamefont{Goniakowski}},
  \bibinfo{author}{\bibfnamefont{F.}~\bibnamefont{Finocchi}}, \bibnamefont{and}
  \bibinfo{author}{\bibfnamefont{C.}~\bibnamefont{Noguera}},
  \bibinfo{journal}{Rep. Prog. Phys.} \textbf{\bibinfo{volume}{71}},
  \bibinfo{pages}{016501} (\bibinfo{year}{2008}).

\bibitem[{\citenamefont{Noguera and Goniakowski}(2008)}]{noguera2008a}
\bibinfo{author}{\bibfnamefont{C.}~\bibnamefont{Noguera}} \bibnamefont{and}
  \bibinfo{author}{\bibfnamefont{J.}~\bibnamefont{Goniakowski}},
  \bibinfo{journal}{J. of Phys: Cond. Matt.}  (\bibinfo{year}{2008}).

\bibitem[{\citenamefont{Gu et~al.}(2009)\citenamefont{Gu, Elfimov, and
  Sawatzky}}]{Gu2009}
\bibinfo{author}{\bibfnamefont{X.}~\bibnamefont{Gu}},
  \bibinfo{author}{\bibfnamefont{I.~S.} \bibnamefont{Elfimov}},
  \bibnamefont{and} \bibinfo{author}{\bibfnamefont{G.~A.}
  \bibnamefont{Sawatzky}} (\bibinfo{year}{2009}),
  \bibinfo{note}{{arXiv:0911.4145v1}}.

\bibitem[{\citenamefont{Ohtomo and Hwang}(2004)}]{ohtomo2004}
\bibinfo{author}{\bibfnamefont{A.}~\bibnamefont{Ohtomo}} \bibnamefont{and}
  \bibinfo{author}{\bibfnamefont{H.~Y.} \bibnamefont{Hwang}},
  \bibinfo{journal}{Nature} \textbf{\bibinfo{volume}{427}},
  \bibinfo{pages}{423} (\bibinfo{year}{2004}).

\bibitem[{\citenamefont{Thiel et~al.}(2006)\citenamefont{Thiel, Hammerl,
  Schmehl, Schneider, and Mannhart}}]{Thiel2006}
\bibinfo{author}{\bibfnamefont{S.}~\bibnamefont{Thiel}},
  \bibinfo{author}{\bibfnamefont{G.}~\bibnamefont{Hammerl}},
  \bibinfo{author}{\bibfnamefont{A.}~\bibnamefont{Schmehl}},
  \bibinfo{author}{\bibfnamefont{C.~W.} \bibnamefont{Schneider}},
  \bibnamefont{and} \bibinfo{author}{\bibfnamefont{J.}~\bibnamefont{Mannhart}},
  \bibinfo{journal}{Science} \textbf{\bibinfo{volume}{313}},
  \bibinfo{pages}{1942} (\bibinfo{year}{2006}).

\bibitem[{\citenamefont{Huijben et~al.}(2006)\citenamefont{Huijben, Rijnders,
  Blank, Bals, {Van Aert}, Verbeeck, {Van Tendeloo}, Brinkman, and
  Hilgenkamp}}]{huijben2006}
\bibinfo{author}{\bibfnamefont{M.}~\bibnamefont{Huijben}},
  \bibinfo{author}{\bibfnamefont{G.}~\bibnamefont{Rijnders}},
  \bibinfo{author}{\bibfnamefont{D.~H.~A.} \bibnamefont{Blank}},
  \bibinfo{author}{\bibfnamefont{S.}~\bibnamefont{Bals}},
  \bibinfo{author}{\bibfnamefont{S.}~\bibnamefont{{Van Aert}}},
  \bibinfo{author}{\bibfnamefont{J.}~\bibnamefont{Verbeeck}},
  \bibinfo{author}{\bibfnamefont{G.}~\bibnamefont{{Van Tendeloo}}},
  \bibinfo{author}{\bibfnamefont{A.}~\bibnamefont{Brinkman}}, \bibnamefont{and}
  \bibinfo{author}{\bibfnamefont{H.}~\bibnamefont{Hilgenkamp}},
  \bibinfo{journal}{Nat. Mater.} \textbf{\bibinfo{volume}{5}},
  \bibinfo{pages}{556} (\bibinfo{year}{2006}).

\bibitem[{\citenamefont{Nakagawa et~al.}(2006)\citenamefont{Nakagawa, Hwang,
  and Muller}}]{nakagawa2006}
\bibinfo{author}{\bibfnamefont{N.}~\bibnamefont{Nakagawa}},
  \bibinfo{author}{\bibfnamefont{H.~Y.} \bibnamefont{Hwang}}, \bibnamefont{and}
  \bibinfo{author}{\bibfnamefont{D.~A.} \bibnamefont{Muller}},
  \bibinfo{journal}{Nature Mater.} \textbf{\bibinfo{volume}{5}},
  \bibinfo{pages}{204} (\bibinfo{year}{2006}).

\bibitem[{\citenamefont{Singh-Bhalla et~al.}(2011)\citenamefont{Singh-Bhalla,
  Bell, andWolter Siemons, Hikita, Salahuddin, Hebard, Hwang, and
  Ramesh}}]{singhbhalla2011}
\bibinfo{author}{\bibfnamefont{G.}~\bibnamefont{Singh-Bhalla}},
  \bibinfo{author}{\bibfnamefont{C.}~\bibnamefont{Bell}},
  \bibinfo{author}{\bibfnamefont{J.~R.} \bibnamefont{andWolter Siemons}},
  \bibinfo{author}{\bibfnamefont{Y.}~\bibnamefont{Hikita}},
  \bibinfo{author}{\bibfnamefont{S.}~\bibnamefont{Salahuddin}},
  \bibinfo{author}{\bibfnamefont{A.~F.} \bibnamefont{Hebard}},
  \bibinfo{author}{\bibfnamefont{H.~Y.} \bibnamefont{Hwang}}, \bibnamefont{and}
  \bibinfo{author}{\bibfnamefont{R.}~\bibnamefont{Ramesh}},
  \bibinfo{journal}{Nat. Phys.} \textbf{\bibinfo{volume}{7}},
  \bibinfo{pages}{80} (\bibinfo{year}{2011}).

\bibitem[{\citenamefont{Xie et~al.}(2011)\citenamefont{Xie, Hikita, Bell, and
  Hwang}}]{xie2011}
\bibinfo{author}{\bibfnamefont{Y.}~\bibnamefont{Xie}},
  \bibinfo{author}{\bibfnamefont{Y.}~\bibnamefont{Hikita}},
  \bibinfo{author}{\bibfnamefont{C.}~\bibnamefont{Bell}}, \bibnamefont{and}
  \bibinfo{author}{\bibfnamefont{H.~Y.} \bibnamefont{Hwang}},
  \bibinfo{journal}{Nat Commun} \textbf{\bibinfo{volume}{2}}
  (\bibinfo{year}{2011}).

\bibitem[{\citenamefont{Segal et~al.}(2009)\citenamefont{Segal, Ngai, Reiner,
  Walker, and Ahn}}]{segal2009}
\bibinfo{author}{\bibfnamefont{Y.}~\bibnamefont{Segal}},
  \bibinfo{author}{\bibfnamefont{J.~H.} \bibnamefont{Ngai}},
  \bibinfo{author}{\bibfnamefont{J.~W.} \bibnamefont{Reiner}},
  \bibinfo{author}{\bibfnamefont{F.~J.} \bibnamefont{Walker}},
  \bibnamefont{and} \bibinfo{author}{\bibfnamefont{C.~H.} \bibnamefont{Ahn}},
  \bibinfo{journal}{Phys. Rev. B} \textbf{\bibinfo{volume}{80}},
  \bibinfo{pages}{241107} (\bibinfo{year}{2009}).

\bibitem[{\citenamefont{Takizawa et~al.}(2011)\citenamefont{Takizawa, Tsuda,
  Susaki, Hwang, and Fujimori}}]{Takizawa2011}
\bibinfo{author}{\bibfnamefont{M.}~\bibnamefont{Takizawa}},
  \bibinfo{author}{\bibfnamefont{S.}~\bibnamefont{Tsuda}},
  \bibinfo{author}{\bibfnamefont{T.}~\bibnamefont{Susaki}},
  \bibinfo{author}{\bibfnamefont{H.}~\bibfnamefont{Y.}~\bibnamefont{Hwang}}, \bibnamefont{and}
  \bibinfo{author}{\bibfnamefont{A.}~\bibnamefont{Fujimori}},
  \bibinfo{journal}{Physical Review B} \textbf{\bibinfo{volume}{84}},
  \bibinfo{pages}{245124} (\bibinfo{year}{2011}).

\bibitem[{\citenamefont{Yu and Zunger}(2014)}]{Yu:2014hx}
\bibinfo{author}{\bibfnamefont{L.}~\bibnamefont{Yu}} \bibnamefont{and}
  \bibinfo{author}{\bibfnamefont{A.}~\bibnamefont{Zunger}},
  \bibinfo{journal}{Nat Commun} \textbf{\bibinfo{volume}{5}},
  \bibinfo{pages}{5118} (\bibinfo{year}{2014}).

\bibitem[{\citenamefont{Kawasaki et~al.}(1994)\citenamefont{Kawasaki,
  Takahashi, Maeda, Tsuchiya, Shinohara, Ishiyama, Yonezawa, Yoshimoto, and
  Koinuma}}]{kawasaki1994}
\bibinfo{author}{\bibfnamefont{M.}~\bibnamefont{Kawasaki}},
  \bibinfo{author}{\bibfnamefont{K.}~\bibnamefont{Takahashi}},
  \bibinfo{author}{\bibfnamefont{T.}~\bibnamefont{Maeda}},
  \bibinfo{author}{\bibfnamefont{R.}~\bibnamefont{Tsuchiya}},
  \bibinfo{author}{\bibfnamefont{M.}~\bibnamefont{Shinohara}},
  \bibinfo{author}{\bibfnamefont{O.}~\bibnamefont{Ishiyama}},
  \bibinfo{author}{\bibfnamefont{T.}~\bibnamefont{Yonezawa}},
  \bibinfo{author}{\bibfnamefont{M.}~\bibnamefont{Yoshimoto}},
  \bibnamefont{and} \bibinfo{author}{\bibfnamefont{H.}~\bibnamefont{Koinuma}},
  \bibinfo{journal}{Science} \textbf{\bibinfo{volume}{266}},
  \bibinfo{pages}{1540} (\bibinfo{year}{1994}).

\bibitem[{\citenamefont{Rijnders et~al.}(2004)\citenamefont{Rijnders, Blank,
  Choi, and Eom}}]{rijnders2004}
\bibinfo{author}{\bibfnamefont{G.}~\bibnamefont{Rijnders}},
  \bibinfo{author}{\bibfnamefont{D.~H.~A.} \bibnamefont{Blank}},
  \bibinfo{author}{\bibfnamefont{J.}~\bibnamefont{Choi}}, \bibnamefont{and}
  \bibinfo{author}{\bibfnamefont{C.-B.} \bibnamefont{Eom}},
  \bibinfo{journal}{Applied Physics Letters} \textbf{\bibinfo{volume}{84}},
  \bibinfo{pages}{505} (\bibinfo{year}{2004}).

\bibitem[{\citenamefont{Yu et~al.}(2012)\citenamefont{Yu, Luo, Yi, Zhang,
  Rossell, Yang, You, Singh-Bhalla, Yang, He et~al.}}]{yu2012}
\bibinfo{author}{\bibfnamefont{P.}~\bibnamefont{Yu}},
  \bibinfo{author}{\bibfnamefont{W.}~\bibnamefont{Luo}},
  \bibinfo{author}{\bibfnamefont{D.}~\bibnamefont{Yi}},
  \bibinfo{author}{\bibfnamefont{J.~X.} \bibnamefont{Zhang}},
  \bibinfo{author}{\bibfnamefont{M.~D.} \bibnamefont{Rossell}},
  \bibinfo{author}{\bibfnamefont{C.-H.} \bibnamefont{Yang}},
  \bibinfo{author}{\bibfnamefont{L.}~\bibnamefont{You}},
  \bibinfo{author}{\bibfnamefont{G.}~\bibnamefont{Singh-Bhalla}},
  \bibinfo{author}{\bibfnamefont{S.~Y.} \bibnamefont{Yang}},
  \bibinfo{author}{\bibfnamefont{Q.}~\bibnamefont{He}}, \bibnamefont{et~al.},
  \bibinfo{journal}{Proc Natl Acad Sci U S A} \textbf{\bibinfo{volume}{109}},
  \bibinfo{pages}{9710} (\bibinfo{year}{2012}).

\bibitem[{\citenamefont{Gozar et~al.}(2007)\citenamefont{Gozar, Logvenov,
  Butko, and Bozovic}}]{gozar2007}
\bibinfo{author}{\bibfnamefont{A.}~\bibnamefont{Gozar}},
  \bibinfo{author}{\bibfnamefont{G.}~\bibnamefont{Logvenov}},
  \bibinfo{author}{\bibfnamefont{V.~Y.} \bibnamefont{Butko}}, \bibnamefont{and}
  \bibinfo{author}{\bibfnamefont{I.}~\bibnamefont{Bozovic}},
  \bibinfo{journal}{Phys. Rev. B} \textbf{\bibinfo{volume}{75}},
  \bibinfo{pages}{201402} (\bibinfo{year}{2007}).

\bibitem[{Kle(2010)}]{Kleibeuker2010}
\bibinfo{journal}{Advanced Functional Materials} \textbf{\bibinfo{volume}{20}},
  \bibinfo{pages}{3490} (\bibinfo{year}{2010}), ISSN \bibinfo{issn}{1616-3028}.

\bibitem[{\citenamefont{Biswas et~al.}(2011)\citenamefont{Biswas, Rossen, Yang,
  Siemons, Jung, Yang, Ramesh, and Jeong}}]{biswas2011}
\bibinfo{author}{\bibfnamefont{A.}~\bibnamefont{Biswas}},
  \bibinfo{author}{\bibfnamefont{P.~B.} \bibnamefont{Rossen}},
  \bibinfo{author}{\bibfnamefont{C.-H.} \bibnamefont{Yang}},
  \bibinfo{author}{\bibfnamefont{W.}~\bibnamefont{Siemons}},
  \bibinfo{author}{\bibfnamefont{M.-H.} \bibnamefont{Jung}},
  \bibinfo{author}{\bibfnamefont{I.~K.} \bibnamefont{Yang}},
  \bibinfo{author}{\bibfnamefont{R.}~\bibnamefont{Ramesh}}, \bibnamefont{and}
  \bibinfo{author}{\bibfnamefont{Y.~H.} \bibnamefont{Jeong}},
  \bibinfo{journal}{Applied Physics Letters} \textbf{\bibinfo{volume}{98}},
  \bibinfo{eid}{051904} (pages~\bibinfo{numpages}{3}) (\bibinfo{year}{2011}).

\bibitem[{\citenamefont{W.~Smekal}(2005)}]{Smekel2005}
\bibinfo{author}{\bibfnamefont{C.~P.} \bibnamefont{W.~Smekal},
  \bibfnamefont{W.S.M.~Werner}}, \bibinfo{journal}{Surf. Interface Anal.}
  \textbf{\bibinfo{volume}{37}}, \bibinfo{pages}{1059} (\bibinfo{year}{2005}).

\bibitem[{\citenamefont{W.S.M.~Werner}(2010)}]{SESSA}
\bibinfo{author}{\bibfnamefont{C.~P.} \bibnamefont{W.S.M.~Werner},
  \bibfnamefont{W.~Smekal}}, \bibinfo{journal}{NIST Database for the Simulation
  of Electron Spectra for Surface Analysis (SESSA) Version 1.3, Standard
  Reference Data Program Database 100
  \textit{http://www.nist.gov/ts/msd/srd/nist100.cfm}}  (\bibinfo{year}{2010}).

\bibitem[{\citenamefont{Pentcheva and Pickett}(2008)}]{pentcheva2008}
\bibinfo{author}{\bibfnamefont{R.}~\bibnamefont{Pentcheva}} \bibnamefont{and}
  \bibinfo{author}{\bibfnamefont{W.~E.} \bibnamefont{Pickett}},
  \bibinfo{journal}{Phys. Rev. B} \textbf{\bibinfo{volume}{78}},
  \bibinfo{pages}{205106} (\bibinfo{year}{2008}).

\bibitem[{\citenamefont{Pentcheva and Pickett}(2009)}]{pentcheva2009}
\bibinfo{author}{\bibfnamefont{R.}~\bibnamefont{Pentcheva}} \bibnamefont{and}
  \bibinfo{author}{\bibfnamefont{W.~E.} \bibnamefont{Pickett}},
  \bibinfo{journal}{Phys. Rev. Lett.} \textbf{\bibinfo{volume}{102}},
  \bibinfo{pages}{107602} (\bibinfo{year}{2009}).

\bibitem[{\citenamefont{Simon et~al.}(2009)\citenamefont{Simon, Zhang, Goodman,
  Xing, Kosel, Fay, and Jena}}]{simon2009}
\bibinfo{author}{\bibfnamefont{J.}~\bibnamefont{Simon}},
  \bibinfo{author}{\bibfnamefont{Z.}~\bibnamefont{Zhang}},
  \bibinfo{author}{\bibfnamefont{K.}~\bibnamefont{Goodman}},
  \bibinfo{author}{\bibfnamefont{H.}~\bibnamefont{Xing}},
  \bibinfo{author}{\bibfnamefont{T.}~\bibnamefont{Kosel}},
  \bibinfo{author}{\bibfnamefont{P.}~\bibnamefont{Fay}}, \bibnamefont{and}
  \bibinfo{author}{\bibfnamefont{D.}~\bibnamefont{Jena}},
  \bibinfo{journal}{Phys. Rev. Lett.} \textbf{\bibinfo{volume}{103}},
  \bibinfo{pages}{026801} (\bibinfo{year}{2009}).

\bibitem[{\citenamefont{Bykhovski et~al.}(1995)\citenamefont{Bykhovski,
  Gelmont, Shur, and Khan}}]{Bykhovski1995}
\bibinfo{author}{\bibfnamefont{A.}~\bibnamefont{Bykhovski}},
  \bibinfo{author}{\bibfnamefont{B.}~\bibnamefont{Gelmont}},
  \bibinfo{author}{\bibfnamefont{M.}~\bibnamefont{Shur}}, \bibnamefont{and}
  \bibinfo{author}{\bibfnamefont{A.}~\bibnamefont{Khan}}, \bibinfo{journal}{J.
  Appl. Phys.} \textbf{\bibinfo{volume}{77}}, \bibinfo{pages}{1616}
  (\bibinfo{year}{1995}).

\bibitem[{\citenamefont{Wetzel et~al.}(2000)\citenamefont{Wetzel, Takeuchi,
  Amano, and Akasaki}}]{wetzel2000}
\bibinfo{author}{\bibfnamefont{C.}~\bibnamefont{Wetzel}},
  \bibinfo{author}{\bibfnamefont{T.}~\bibnamefont{Takeuchi}},
  \bibinfo{author}{\bibfnamefont{H.}~\bibnamefont{Amano}}, \bibnamefont{and}
  \bibinfo{author}{\bibfnamefont{I.}~\bibnamefont{Akasaki}},
  \bibinfo{journal}{Phys. Rev. B} \textbf{\bibinfo{volume}{61}},
  \bibinfo{pages}{2159} (\bibinfo{year}{2000}).

\bibitem[{\citenamefont{Cancellieri et~al.}(2011)\citenamefont{Cancellieri,
  Fontaine, Gariglio, Reyren, Caviglia, F\^ete, Leake, Pauli, Willmott, Stengel
  et~al.}}]{Cancellieri2011}
\bibinfo{author}{\bibfnamefont{C.}~\bibnamefont{Cancellieri}},
  \bibinfo{author}{\bibfnamefont{D.}~\bibnamefont{Fontaine}},
  \bibinfo{author}{\bibfnamefont{S.}~\bibnamefont{Gariglio}},
  \bibinfo{author}{\bibfnamefont{N.}~\bibnamefont{Reyren}},
  \bibinfo{author}{\bibfnamefont{A.~D.} \bibnamefont{Caviglia}},
  \bibinfo{author}{\bibfnamefont{A.}~\bibnamefont{F\^ete}},
  \bibinfo{author}{\bibfnamefont{S.~J.} \bibnamefont{Leake}},
  \bibinfo{author}{\bibfnamefont{S.~A.} \bibnamefont{Pauli}},
  \bibinfo{author}{\bibfnamefont{P.~R.} \bibnamefont{Willmott}},
  \bibinfo{author}{\bibfnamefont{M.}~\bibnamefont{Stengel}},
  \bibnamefont{et~al.}, \bibinfo{journal}{Phys. Rev. Lett.}
  \textbf{\bibinfo{volume}{107}}, \bibinfo{pages}{056102}
  (\bibinfo{year}{2011}).

\end{thebibliography}
\end{document}